\documentstyle[12pt]{article}
 \textwidth
 16.4cm
 \oddsidemargin
 2.5cm
 \advance\oddsidemargin
 by
 -1in
 \evensidemargin
 0.0cm
 \advance\evensidemargin
 by
 -1in
 \marginparwidth
 1.9cm
 \marginparsep
 0.4cm
 \marginparpush
 0.4cm
 \topmargin
 -1.5cm
 \advance\topmargin
 by
 -0.0in
 \textheight
 23.2cm
 \makeindex

 \newcommand\la{\langle}
 \newcommand\ra{\rangle}
 \newcommand\noi{\noindent}
 \newcommand\beq{\begin{equation}}
 \newcommand\eeq{\end{equation}}
 \newcommand\beqn{\begin{eqnarray}}
 \newcommand\eeqn{\end{eqnarray}}

\begin{document}
\vspace*{3cm}

\centerline{\Large
\bf 
Charmonium and lepton pair production }

\medskip

\centerline{\Large
{\bf with medium energy antiprotons}
\footnote{The talk given at the meeting {\sl ''Perspectives of
Hadron Physics''}, GSI, Darmshtadt, June 8, 1998}
}

\vspace{.5cm}

\begin{center}
 {\large
Boris~Kopeliovich}\\
\medskip

{\sl
Max-Planck
Institut
f\"ur
Kernphysik,
Postfach
103980,
69029
Heidelberg,
Germany}\\

{\sl
$\&$ Joint
Institute
for
Nuclear
Research,
Dubna,
141980
Moscow
Region,
Russia}

\end{center}

\vspace{.5cm}
\begin{abstract}

The medium energy antiproton beam facility which is under discussion
at GSI offers new opportunities to study the QCD of heavy flavours,
nuclear structure and exotic mesons. We discuss these topics and briefly 
highlight other problems in charm production and proton structure.

\end{abstract}

\bigskip



\section{Charmonium production off nuclei}
\subsection{Properties of charmonia}
\medskip

Charmonium spectroscopy is known as a laboratory
for heavy quark QCD.
Realistic potential models \cite{buch} successfully describe
the mass spectrum. Not much is known, however, about charmonium
properties. Even the cross section of interaction with
a nucleon, $\sigma_{tot}(\Psi N)$,  is not measured. 
One desperately needs to know
it in order to interpret the data on charmonium production in 
relativistic heavy ion collisions.

The sources of information about $\sigma_{tot}(J/\Psi N)$
are the photoproduction data on a proton and hadroproduction off nuclei
\cite{hk}.
The former, analysed with VDM gives $\sigma_{tot}(J/\Psi N)=1.2\,mb$
at the SPS energy $\sqrt{s_{\Psi}}
=10\,GeV$, while the latter
analysed within the optical model results in
$\sigma_{tot}(J/\Psi N)\approx 6\,mb$.
Data for $\Psi'(2S)$ production in both cases 
also lead to a puzzling conclusion that
$\sigma_{tot}(\Psi' N)\approx \sigma_{tot}(J/\Psi N)$
contradicting the expectation that $\Psi'$
having about twice a big radius as $J/\Psi$ \cite{buch} 
has to interact much stronger.
These controversies probably mean that the one-channel
approach fails and the higher charmonium excitations must 
come to play \cite{hk}.
The key parameters which control the solution of the
two-channel problem for charmonium production off nuclei are
\beqn
\epsilon &=& 
\frac{\la\Psi'|\hat\sigma|J/\Psi\ra}
{\la J/\Psi|\hat\sigma|J/\Psi\ra}\\
r &=&
\frac{\la\Psi'|\hat\sigma|\Psi'\ra}
{\la J/\Psi|\hat\sigma|J/\Psi\ra}\\
R &=&
\frac{\sigma(hN\to \Psi'X)}
{\sigma(hN\to J/\Psi X)}
\label{1.1}
\eeqn

One can only roughly estimate these parameters from
data on charmonium hadroproduction off nuclei
since the production dynamics is poorly known.
A better information could be extracted from virtual 
photoproduction, but no data are available and are
not expected in the near future.

Direct production $\bar pp \to J/\Psi(\Psi')$ has an advantage
of known dynamics. besides, it provides the 
lowest energy of produced charmonium,
$p_L(J/\Psi)=4.06\,GeV$, $p_L(\Psi')=6.2\,GeV$\footnote{this is much lower
energy than can be achieved in practice in experiments with inverse
kinematics at SPS. Even at $x_F=-0.5$ the $J/\Psi$ momentum is $9\,GeV$.}.
At this energy the contribution from other channels is tiny and
one can reliably measure the $\Psi$-nucleon absorption cross section.

The production cross section off nuclei at the energy of the peak
is, however, strongly suppressed compared
to $\bar pp\to \Psi$ due to Fermi motion by a factor \cite{fs},
\beq
\frac{\sigma(\bar pA\to\Psi A')}
{Z\,\sigma(\bar pp\to\Psi)} = 
\pi\,\frac{M_N}{M_{\Psi}}\ 
\frac{\Gamma_{\Psi}}
{k_F}\ ,
\label{1.2}
\eeq
\noi
where $k_F$ is the mean value of the Fermi momentum.
If the beam momentum resolution $\Delta p_b$ 
exceeds $\Gamma_{\Psi}$
(most probably), then $\Gamma_{\Psi}$ should be replace by
$\Delta p_b$ \cite{fs} what may substantially enhance the ratio in (\ref{1.2}). 
This does not mean, however, that with
a worse resolution $\Delta p_b$ the nuclear cross section is larger.
The ratio (\ref{1.2}) grows just due to 
decreasing denominator $\sigma(\bar pp\to\Psi)$.
At $\Delta p_b = 1\,MeV$ the production  rate 
off nuclei is suppressed by three orders of magnitude. 

The dependence of $J/\Psi$ and $\Psi'$ production cross 
section on the beam momentum can be evaluated using the 
Gaussian form for the Fermi momentum distribution with 
$k_F=200\,MeV$. Fig.~1 shows such a dependence in arbitrary units,
however, the relative $\Psi'$ to $J/\Psi$ production rate is fixed by
\beq
\frac{N_{\Psi'}}{N_{J/\Psi}} = 
\frac{\Gamma^{\bar pp}_{\Psi'}}
{\Gamma^{\bar pp}_{J/\Psi}}\,
e^{-\Delta\sigma\,\la T\ra}\ ,
\label{1.3}
\eeq
\noi
where $\la T\ra$ is the mean value of the nuclear thickness function,
$\Delta\sigma=\sigma_{tot}(\Psi' N)-\sigma_{tot}(J/\Psi N)
\approx 9\,mb$.
\begin{figure}[tbh]
\includegraphics{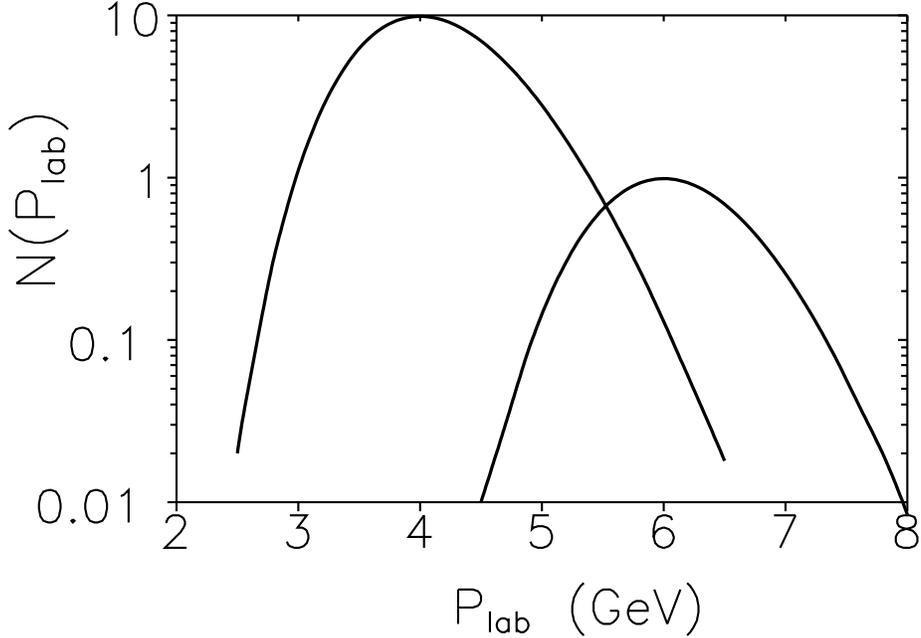}
\begin{center}
\vspace{8cm}
\parbox{13cm}
{\caption[Delta]
{\sl Production rate of $J/\Psi$ and $\Psi'$ in $\bar p\,Pb$
annihilation as a function of the beam momentum.}}
\end{center}
\end{figure}

Fig.~1 shows that in the beam momentum range 
$p_L = 5 - 6\ GeV$ one can produce 
either $J/\Psi$ or $\Psi'$ at the same beam momentum.
As a result, these two states strongly interfere, leading to
a very much modified transparency for the charmonia.
The value of the parameter $R$ varies with $p_L$ from 
zero to infinity, and
either $J/\Psi$ or $\Psi'$ can be enhanced by the nucleus, 
rather than suppressed. Indeed, the effective ''absorption'' 
cross section may be negative dependent on $R$ \cite{prl},
since $\epsilon < 0$,
\beqn
\sigma^{J/\Psi}_{abs}&=&
\sigma_{tot}(J/\Psi\,N)\,
\left[1+\epsilon\,R\,F_A^2(q_f)\right]<0\ 
\ \ if\ R\gg 1\nonumber\\
\sigma^{\Psi'}_{abs}&=&
\sigma_{tot}(\Psi'\,N)\,
\left[1+\frac{\epsilon}{r\,R}\,F_A^2(q_f)\right]<0\
\ \ if\ R\ll 1\ .
\label{1.4}
\eeqn
\noi
Here the nuclear form factor depends on the longitudinal momentum
transfer related to the formation length \cite{prl},
\beq
q_f=\frac{M_{\Psi'}^2-M_{J/\Psi}^2}{2\,p_L}\ .
\label{1.5}
\eeq

Note that even if data on virtual photoproduction of charmonia were 
available, they would have quite a restricted range of variation 
of the parameter $0.5 < R < 1$ corresponding to
$Q^2=0$ and $Q^2\to\infty$, respectively.

\subsection{\boldmath$J/\Psi$ probe for nuclear structure} 

When the beam momentum is far from the overlap region in Fig.~1
the interference effects vanish and the kinematics is
certain. This makes $J/\Psi$ production a perfect tool to study
Fermi distribution of bound protons, especially its high-momentum tail.
Recently experiment E850 at BNL \cite{heppel}
on $A(p,2p)A'$ with detection of the recoil
neutron found a strong correlation between the disbalance of the 
transverse momenta of the protons and the neutron momentum. This is the 
first solid evidence in favour of hypothesis that the high-momentum tail of
Fermi distribution is due to two-nucleon correlations, rather than
to the mean nuclear field.

Such a reaction, however, is unable to measure the longitudinal component
of the Fermi momentum because of color transparency effect \cite{jk}.
On the other hand, deep-inelastic lepton scattering is
sensitive only to the longitudinal component. $J/\Psi$ production
in $\bar p$ annihilation allows to measure both $k_L$ and
$k_T$ with a high precision, provided that the beam momentum is known with
a high resolution.

The correlation between the longitudinal and transverse components 
of the Fermi momentum dictated by the light-cone nuclear dynamics
is quite different from the intuitive expectation: $\la k_T^2\ra$
grows with $k_L$. Indeed, a bound nucleon carries a fraction of the 
total nuclear light-cone momentum, $\alpha = 1/A + \Delta_{\alpha}$,
where $\Delta_{\alpha}$ is related to longitudinal Fermi motion.
The energy denominator has a form,
\beq
\frac{k_T^2}{\alpha(1-\alpha)} +
\frac{M_N^2}{\alpha} + (A-1)^2\,\frac{M_N^2}{\alpha}
- A^2\,M_N^2 \nonumber\\
= \frac{k_T^2+A^2\,M_N^2\,\Delta_{\alpha}^2}
{\alpha(1-\alpha)}\ .
\label{1.6}
\eeq
\noi
Therefore, 
\beq
\la k_T^2\ra = A^2\,M_N^2\,\Delta_{\alpha}^2 = k_L^2\ .
\label{1.7}
\eeq
\noi
Thus, we arrived at quite an unusual correlation: the mean value
of $k_T^2$ grows with $k_L$. 
Note that this correlation is independent of the origin
of high Fermi momenta, mean field or two-nucleon correlation.
To disentangle between them in $J/\Psi$ production by antiprotons 
one should detect a recoil neutron, like is done in \cite{heppel}.

\section{Search for the 4-quark mesons in Drell-Yan reaction}

A diquark, {\it i.e.} a $qq$ pair in an antitriplet color state is different
from $\bar q$ only by spin and mass, unless the interaction
is hard enough to resolve its $qq$ structure.
Replacing the $\bar q$ in a meson by a diquark one gets a baryon.
It looks very natural to replace the quark in a baryon by antidiquark.
Then one arrives at a 4-quark meson having a diquark-antidiquark
structure, $D-\bar D$. One of the candidates for such a state is
the $C(1480)$ meson, found in the OZI suppressed decay mode
$\phi\pi^0$ \cite{land} (see also in \cite{kp}).

A $\bar pp$ collision is a natural source for such
4-quark mesons, since the diquark and antidiquark already pre-exist.
One should annihilate the beam-target valence $\bar q-q$ pair and the rest 
is an excited $\bar DD$ pair, as is shown in Fig.~2a.
\begin{figure}[tbh]
\includegraphics{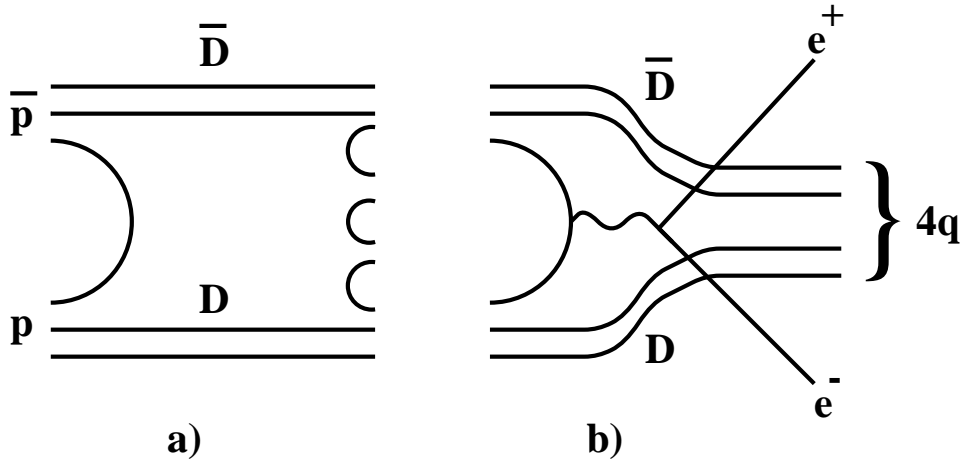}
\begin{center}
\vspace{5.9cm}
\parbox{13cm}
{\caption[Delta]
{\sl {\bf a)} $\bar pp$ interaction with annihilation of a pair of
projectile valence $\bar qq$ pair. The produced diquark-antidiquark
state mass is $M_{\bar DD}>2M_N$.
{\bf b)} The same as in fig.~2a, but accompanied with
production of a Drell-Yan lepton pair. A 4-quark 
state of any mass can be produced.}}
\end{center}
\end{figure}
The effective mass of $\bar DD$ is, however, larger than
$2M_N$ and a decay to a lighter 4-q state is difficult. 
Indeed, decay of the color string as it is shown in Fig.~2a
leads to a production of a baryon-antibaryon pair and mesons.

An effective way to form a desired 4-quark state is to
produce a lepton pair via Drell-Yan mechanism as is shown in 
Fig.~2b. First of all, the lepton pair signals that the $\bar qq$
annihilation did occur. Secondly, it takes away an extra energy
and the rest 4-quark state may have any mass. Thirdly,
the lepton pair is easy to detect and it fixes the missing mass 
of the 4-quark system, provided that the lepton momenta are well 
measured. One can also detect the decay products of the 
4-quark meson which would help much in establishing its 
quantum numbers.

This reaction suggested in \cite{dubna} 
can be studied complementary to charmonium production.
The preferred effective mass interval is $1.5 < M_{e^+e^-}
< 3\,GeV$. The reaction was simulated in \cite{dubna} for 
kinematics corresponding to the geometry of the SuperLEAR
internal beam detector, and the cross section parameterized as,
\beq
\frac{d\sigma(\bar pp\to e^+e^-\,X)}
{dM_{e^+e^-}} = 
\frac{0.21}{M^3_{e^+e^-}}\,
e^{-11.7\,\tau}\ (\mu b)\ ,
\label{2.1}
\eeq
\noi
where $\tau=M^2_{e^+e^-}/s$ and $M_{e^+e^-}$ is in $GeV$.

It was found in \cite{dubna} that with luminosity $10^{32}\,cm^{-2}
sec^{-1}$, the overall efficiency $0.5$ and the trigger efficiency
$0.7$ one can expect $\sim 1200$ events a day with a $12\,GeV$ antiproton
beam. For a light 4-quark system, $0.9 < M_{4q} < 2\ GeV$, the expected 
rate is $\sim 300$ events a day. For this mass interval the production
rate is maximal for the beam momentum range $5-8\, GeV$.

\section{What else?}

\subsection{Open charm production}

\begin{itemize}

\item {\sl Inclusive production of leading $D$ and $\bar D$ mesons}.
$D$($\bar D$) mesons (not to be mixed up with diquarks) 
are produced in $\bar pp \to D(\bar D) X$ at high $x_F$
in a very unusual configuration: the
major fraction of the longitudinal momentum is carried by the
projectile light quark. In this configuration the $D$ meson is effectively
stopped by interaction with nuclear matter, because after the $D$ meson 
breaks up the slow $c$ quark produces a $D$ meson with low $x_F$.
This is very different from
propagation of a formed $D$ meson containing a leading $c$ quark, 
which loses only a tiny fraction of its
momentum interacting with medium. Nuclear suppression of $D$ meson
production rate versus Feynman $x_F$ would provide precious information about
space-time pattern of formation of the $D$ meson wave function.

\item {\sl Exclusive production $\bar pp \to D\bar D$}.
In this case a very strong leading effect is expected:
the $D$($\bar D$) mesons are predominantly produced in
the fragmentation regions of $\bar p$($p$).
When this reaction occurs in a nuclear environment, nuclear suppression of 
production rate serves as a direct measurement of $D$ and $\bar D$
absorption via inelastic interactions with bound nucleons
on the way out of the nucleus. As different from inclusive $D$ meson
production, the $DN$ and $\bar D N$ inelastic cross sections can be 
measured.

\end{itemize}

\subsection{Proton structure and nature of baryon number}

\begin{itemize}

\item {\sl Quark intrinsic momentum in a diquark}.
Comparative measurement of transverse momenta of pions produced in
annihilation and non-annihilation channels of $\bar pp$ interaction
provides unambiguous information about the diquark size.
In the former case the (anti)diquark in the (anti)proton breaks up and releases
the intrinsic quark momenta, which are expected to be 
much larger than what we usually
observe in inelastic hadronic collisions. Indeed, 
there are many indications \cite{mauro}, 
both experimental and theoretical, that 
the diquark in a proton has a size much smaller
than the proton radius. We suggest to use this 
as a very sensitive way
to measure the quark Fermi-momentum inside the
diquark.
A strong $u-d$ quark correlation is expected due to the 
gluonic structure of the QCD
vacuum (instanton model \cite{instanton}) and even 
in potential models \cite{johan}.

\item {\sl Mechanisms of baryon stopping} is a hot topic in physics of
relativistic heavy ion collisions \cite{ck}. One can study this in more detail in
$\bar pp$ annihilation where the baryon number flows over the whole
rapidity interval.
According to the modern classification \cite{rv,kz6} 
baryon number can be transferred
by the diquark (important only at rather low energy, up to a few GeV \cite{k}),
by a valence quark plus string junction \cite{kz3} 
(mid energies) and by a pure
gluon field \cite{k,kz1,kz2,kz4} (=string junction). 
These mechanisms have a very different
multiplicity of produced particles, 1:2:3, respectively.
One can effectively disentangle these three mechanisms by analyzing
the multiplicity distribution in annihilation events \cite{kz2}. 
One can also test the so-called Eylon hypothesis \cite{eh,kz6} 
that annihilation is related through the
unitarity relation to the Pomeron, i.e. it does not contribute to the
$\bar pp$ and $pp$ total cross section difference.

\item Antiproton production with high $p_T$ in $\bar pp$ interaction is known
to be due to high-$p_T$ scattering of the projectile antidiquark.
Therefore, a high-$p_T$ antiproton produced in $\bar p\,A$ collision
should attenuate due to annihilation of the diquark propagating
through nuclear matter \cite{kz7}. 
The annihilation (break up) cross section strongly
depends on the diquark size (an analogue to color  transparency). Therefore one
can get precious information about the diquark radius \cite{kz7}.

\end{itemize}

{\sl Summarizing}, a high luminosity 
antiproton beam at GSI would open new opportunities
for study of QCD dynamics of strong 
interactions and hadronic structure.
Many of them are unique.
This talk covers only a small part of the possibilities particularly presented 
by other speakers at this meeting. 

\bigskip

{\bf Acknowledgements:} I am grateful to J\"org H\"ufner and
Wolfram Weise for collaboration in formulation of 
the problems presented in
the first two sections. The ideas proposed for experimental study 
in the last section were generated during stimulating discussion with
A.~Gillitzer, S.~Paul and W.~Weise. 
I am also thankful to Sverker Fredriksson who read the paper 
and made valuable comments and Josef Pochodzalla
for useful discussions.

\end{document}